\documentstyle[12pt,epsfig,amssymb]{article}
\begin{document}
\begin{flushright}
IHEP-01-30\\
\end{flushright}

\begin{center}

{\Large \bf Impact of the three-loop corrections on the QCD analysis of   
the deep-inelastic-scattering data}

\vspace{1cm}
{\bf S.~I.~Alekhin}

\vspace{0.1in}
{\baselineskip=14pt Institute for High Energy Physics, 142281 Protvino, Russia}

\begin{abstract}
We perform the analysis of the existing inclusive deep inelastic scattering 
(DIS) data within NNLO QCD approximation. The parton distributions 
functions (PDFs) and the 
value of strong coupling constant 
$\alpha_{\rm s}(M_{\rm Z})=0.1143\pm0.0013~({\rm exp})$ are obtained. 
The sensitivity of the PDFs to the uncertainty in the value 
of the NNLO corrections to the 
splitting functions is analyzed. It is shown that the PDFs errors due to 
this uncertainty 
is generally less than the experimental uncertainty in PDFs
through the region of $x$ spanned by the existing DIS data.
\end{abstract}
\end{center}
{\bf PACS numbers:} 13.60.Hb,06.20.Jr,12.38.Bx\\
{\bf Keywords:} deep inelastic scattering, higher-order corrections

\newpage
{\bf 1.} The account of higher-order corrections in an analysis
based on the QCD perturbative expansions is very important. For the 
relevant processes measured to the moment the typical value of the 
strong coupling constant $\alpha_{\rm s}$ is $O(0.1)$ and
the convergence of series in $\alpha_{s}$ is slow. For  
the deep-inelastic-scattering (DIS) process this problem is especially 
important for the largest and the lowest $x$ regions, where 
the coefficients of the series contain the terms proportional to 
``large logarithms''. Meanwhile due to great technical 
difficulties the progress in calculation
of the higher-order QCD corrections is not so fast. In particular 
for the case of the inclusive DIS structure functions only the two-loop 
QCD corrections have been calculated completely \cite{Furmanski:1982cw}. 
The three-loop (NNLO) case coefficient functions are known exactly
\cite{SanchezGuillen:1991iq}, while for the corrections to the
splitting functions only the even Mellin moments up to 8 and 
some asymptotes were known to the recent time 
\cite{Larin:1997wd,Gracey:1994nn}.

An attempt to combine all available information about splitting functions
in order to obtain reasonable approximation to the exact expressions
was done in Refs.~\cite{vanNeerven:2000ca,vanNeerven:2000ca1}. 
The result of this study is 
the set of approximate NNLO splitting functions in the $x$ space
supplied by the estimate of their possible variation due to effect of 
the highest moments. These approximate splitting functions have been used 
in the analysis of
Ref.~\cite{Martin:2000gq} aimed to estimate the effect of the NNLO
QCD corrections on the shape of the parton distributions functions (PDFs)
extracted from the global fit. 
Meanwhile the gluon distribution obtained in this analysis 
turned out to be sensitive to the uncertainties of the  
NNLO splitting functions given in
Refs.~\cite{vanNeerven:2000ca,vanNeerven:2000ca1}.
In particular at $x \sim 10^{-4}$ and  
$Q^2=20~$GeV$^2$ the error on the gluon distribution due to this 
uncertainty is 
about 35\%, which is much larger than the experimental error on the 
gluon distribution obtained in the two-loop 
analysis of the existing DIS data \cite{Alekhin:2001ch}.

Fortunately the Mellin moments 
of the splitting functions up to 12 were calculated 
recently in Ref.~\cite{Retey:2000nq} that allowed to elaborate new set of 
the approximate splitting functions with much narrower uncertainty range
\cite{vanNeerven:2000wp}.
In this paper we describe the results of our 
analysis of the existing DIS data with account of the NNLO QCD 
corrections. The analysis is based on  the recent splitting functions 
given in Ref.~\cite{vanNeerven:2000wp}. Our main aim is 
to study the effect of NNLO corrections on the PDFs 
and the value of $\alpha_s$ extracted from the data with a 
particular attention paid on the errors due to the 
remaining uncertainty of the NNLO splitting functions.

\begin{figure}
\centerline{\epsfig{file=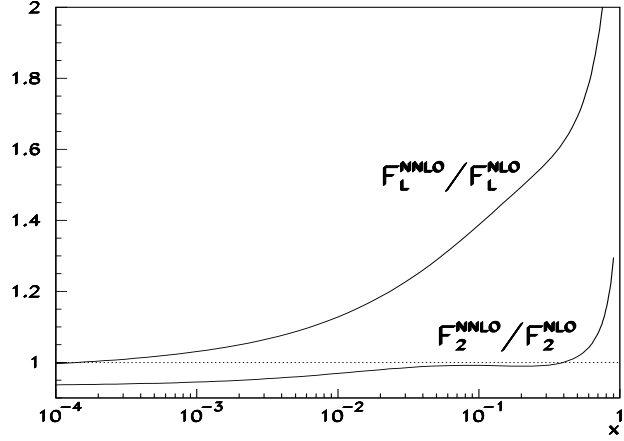,width=10cm,height=8cm}}
\caption{The ratios of the leading twist
structure functions $F_{2,\rm L}$
calculated in the NNLO and the NLO approximations.}
\label{fig:bc2}
\end{figure}

\begin{figure}
\centerline{\epsfig{file=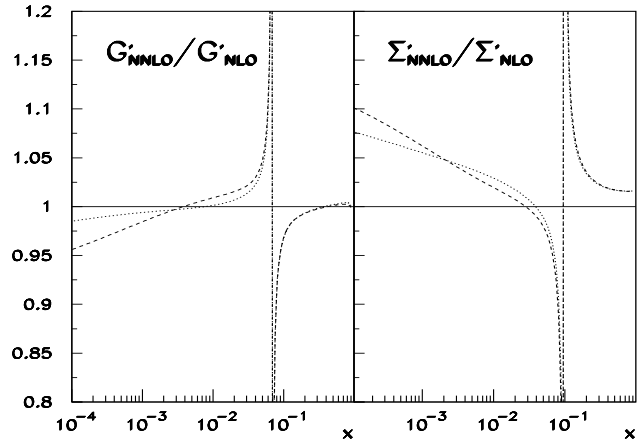,width=10cm,height=8cm}}
\caption{Ratio of the logarithmic derivatives of the gluon 
distribution $G'=d\ln G/d \ln Q$
calculated in the NNLO and the NLO approximations (left);
the same for the singlet distribution (right). The dotted curves 
correspond to the choice A and the dashed curves -- to the choice B
for the splitting functions approximations of 
Ref.~\protect\cite{vanNeerven:2000wp}.}
\label{fig:bp2}
\end{figure}

{\bf 2.} Our theoretical ansatz and the fitting procedure 
are the same as in our previous analysis of Ref.~\cite{Alekhin:2001ch},
except that now we use the NNLO QCD approximation both for the splitting
functions and the leading twist (LT)
coefficient functions of DIS. Impact of the NNLO 
corrections to the coefficient functions on the 
values of $F_2$ and $F_{\rm L}$ is illustrated in 
Fig.\ref{fig:bc2}, where we give the ratios of these structure  
functions in the NNLO approximation to the ones 
in the NLO approximation. The NNLO contributions 
to the coefficient functions in the form 
given in Ref.~\cite{vanNeerven:2000ca}
were used in the calculations. The input for both NLO and NNLO 
calculations was chosen the same as in Ref.~\cite{vanNeerven:2000ca}:
The gluon  distribution $xG(x)=x^{-0.37}(1-x)^5$, 
the total singlet distribution  
$x\Sigma(x)=0.6x^{-0.3}(1-x)^{3.5}(1+5x^{0.8})$, 
the number of flavors $N_f=4$,   
and the value of $\alpha_{\rm s}=0.2$. 
The largest effect of the NNLO corrections to the 
coefficient functions on the values of $F_{\rm 2,L}$
is the rise of $F_{\rm L}$ at large
$x$. Nevertheless for the analysis of existing data this rise
is not so important due to sensitivity of the data to variation 
of $F_{\rm L}$ at large $x$ is rather poor.
Much more important is suppression of the structure 
function $F_2$ by $\sim 5$\% at small $x$
since the precision of existing data on $F_2$ is 
$O(1\%)$ in this region. 
Effect of the NNLO corrections on the splitting functions
is demonstrated in Fig.\ref{fig:bp2}, where the 
ratios of the logarithmic derivatives 
of the gluon and the singlet distributions calculated in the NNLO 
and the NLO approximations are plotted. We used 
in the calculations the approximations of the splitting functions
from Ref.~\cite{vanNeerven:2000wp}
and the input distributions from Ref.~\cite{vanNeerven:2000ca}.
Different curves correspond to the two choices of the splitting 
functions which give the range of the uncertainty of the latter.
One can see that the NNLO corrections to the splitting functions
change the ``speed'' of evolution moderately: At the scale of $Q\sim 10~$GeV
the derivatives change by $\lesssim10\%$ at smallest $x$ and even 
less at the largest $x$ in question (the spike at $x\sim0.1$ is just due 
to the QCD evolution has crossover point here and the 
derivatives are very small in this region).
As a result the main effect of the NNLO corrections
is due to corrections to the coefficient function for $F_2$. 
Figs.\ref{fig:bc2},\ref{fig:bp2} may be used 
for the benchmark of our NNLO evolution code as well. For this purpose one 
can compare these figures with  Fig.10 of 
Ref.~\cite{vanNeerven:2000ca} and Fig.4 of Ref.~\cite{vanNeerven:2000wp}
correspondingly and convince that the agreement of both codes
is perfect\footnote{The extensive cross-check of the different 
NNLO evolution codes is underway now and the results will be released 
at the WWW page of the Les Houches workshop ``Physics at TeV colliders''
(http://pdf.fnal.gov/LesHouches.htm).}.

\begin{table}
\caption{The numbers of data points (NDP) and the $\chi^2$ values 
for the separate experimental data sets used in the analysis.}
\begin{center}
\begin{tabular}{cccc} 
&\multicolumn{2}{c}{NDP}& \\ \cline{2-3}
Experiment&proton&deuterium&$\chi^2$/NDP \\  \hline
SLAC-E-49A  &58 &58   &0.64   \\  
SLAC-E-49B  &144&135 &1.35     \\  
SLAC-E-87   &90&90   &1.07   \\  
SLAC-E-89A   &66&59  &1.46    \\  
SLAC-E-89B  &79&62   &1.12   \\  
SLAC-E-139   &--&16  &0.57    \\  
SLAC-E-140  &--&26   &0.90   \\  
BCDMS  &351&254 &1.17        \\  
NMC    &245&245 &1.29   \\   
H1(96-97)    &122&-- &1.12   \\   
ZEUS(96-97)    &161&-- &1.16   \\   \hline
TOTAL  &1316&945 &1.13     \\ 
\end{tabular}
\end{center}
\label{tab:ndp}
\end{table}

The boundary LT PDFs fitted to the data were parameterized within 
the scheme with fixed number of flavors at $N_f=3$.
At our starting value of the QCD evolution $Q_0^2=9$~GeV$^2$ they read
$$
xu_{\rm V}(x,Q_0)=\frac{2}{N^{\rm V}_{\rm u}}
x^{a_{\rm u}}(1-x)^{b_{\rm u}}(1+\gamma_2^{\rm u}x),~~
xd_{\rm V}(x,Q_0)=\frac{1}{N^{\rm V}_{\rm d}}x^{a_{\rm d}}(1-x)^{b_{\rm d}},
$$
$$
xu_{\rm S}(x,Q_0)=\frac{A_{\rm S}}{N_{\rm S}}
\eta_{\rm u} x^{a_{\rm su}}(1-x)^{b_{\rm su}},~~
xd_{\rm S}(x,Q_0)=\frac{A_{\rm S}}{N^{\rm S}}
x^{a_{\rm sd}}(1-x)^{b_{\rm sd}},~~
xs_{\rm S}(x,Q_0)=\frac{A_{\rm S}}{N^{\rm S}}\eta_{\rm s}
x^{a_{\rm ss}}(1-x)^{b_{\rm ss}},
$$
\begin{equation}
xG(x,Q_0)=A_{\rm G}x^{a_{\rm G}}(1-x)^{b_{\rm G}}
(1+\gamma^{\rm G}_1\sqrt{x}+\gamma^{\rm G}_2 x),
\end{equation}
where $u,d,s,G$ are the up, down, strange quarks,
and gluons distributions respectively; 
the indices $V$ and $S$ correspond to the valence 
and sea quarks. The parameters $N^{\rm V}_{\rm u}, N^{\rm V}_{\rm d}$  and 
$A_{\rm G}$ were not fitted, instead they were calculated 
from the other parameters using the conservation of the partons momentum 
and the fermion number. The normalization 
parameter $N^{\rm S}$ is also calculated from the other parameters 
in such way that the normalization parameter 
$A_{\rm s}$ correspond to the total momentum carried
by the sea quarks. The parameter $\eta_{\rm s}$ was fixed at 0.42 and 
the other sea distributions parameters
were constrained as $a_{\rm su}=a_{\rm sd}=a_{\rm ss}$, 
$b_{\rm ss}=(b_{\rm su}+b_{\rm sd})/2$. 

\begin{figure}
\centerline{\epsfig{file=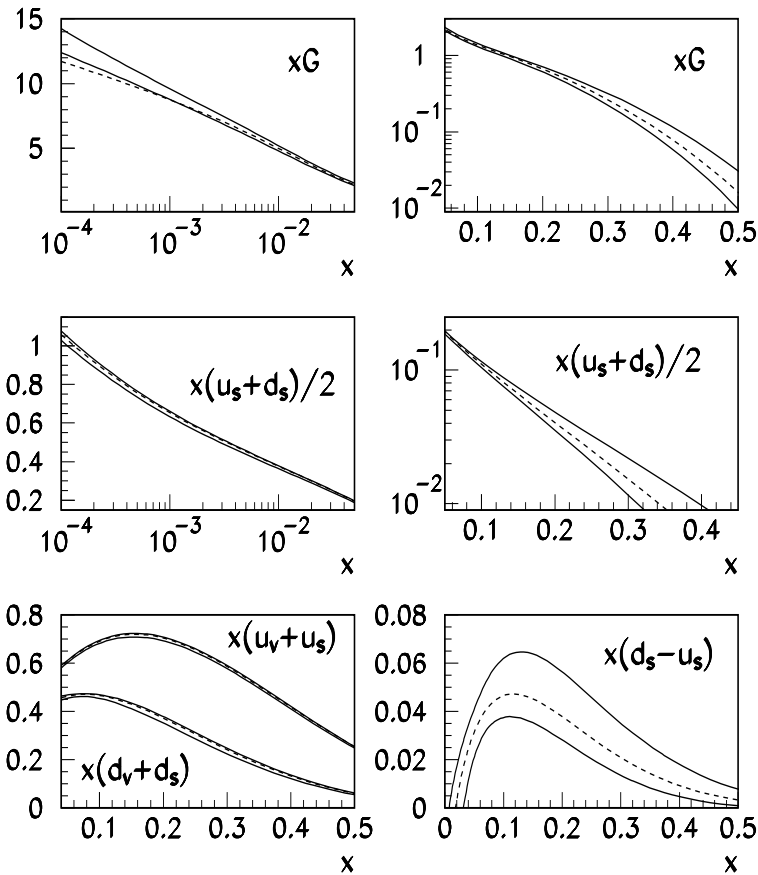,width=12cm,height=14cm}}
\caption{The selected PDFs obtained from the 
NNLO fits with the different choice of the NNLO splitting functions.
(Full curves: the $1\sigma$ experimental bands for the fit 
with the splitting functions chosen as average of the 
variants A and B of Ref.~\protect\cite{vanNeerven:2000wp};
dashed curves: the central values for the fit with variant B
of Ref.~\protect\cite{vanNeerven:2000wp}).}
\label{fig:pdf}
\end{figure}

The LT structure functions $F_{\rm 2,L}$ obtained using 
these PDFs evolved using the GLAPD equations \cite{Gribov:1972ri}
were corrected for  
the target-mass correction and the high-twist contribution as well as it was 
done in our earlier analysis 
of Ref.~\cite{Alekhin:2001ch}. The result  
was substituted to the regular expression for the 
inclusive DIS cross section, which was fitted to the 
existing data 
varying parameters of PDFs, the value of $\alpha_{\rm s}$, and the 
high-twist contributions to $F_{\rm 2,L}$.
For our nominal fit we use the NNLO splitting functions 
obtained as the average of the variants A and B
given in Ref.~\cite{vanNeerven:2000wp}.

We used for the analysis the data on DIS of charged leptons off 
the proton and deuterium targets.
The data set coincides in part with the one used in 
Ref.~\cite{Alekhin:2001ch}. The difference from that analysis    
is that now we include in the fit the H1 data of Ref.~\cite{Adloff:2000qk}
and the ZEUS data of Ref.~\cite{Chekanov:2001qu}
collected in 1996-97 instead of earlier 
data of these collaborations. Besides, we drop 
the data from the 
FNAL-E-665 experiment \cite{Adams:1996gu} since they have no 
impact on the results of the analysis due to large experimental errors.
It was also checked that inclusion of the high $Q^2$ data of
Ref.~\cite{Adloff:2000ah} does not decrease 
the experimental errors in the fitted
values of PDFs and $\alpha_{\rm s}$ and for this reason they were not
included in our analysis. The same is valid for the ZEUS data 
of Ref.~\cite{Chekanov:2001qu} with $Q^2~>~300$~GeV$^2$ and we 
also discarded these data points from the analyzed data set
in order to escape the region where the corrections due to the 
$Z$-boson contribution should be taken into account.
The data points with $Q^2<2.5~$GeV$^2$ and $x>0.75$ were cut
in order to improve the perturbative stability of the results and 
to minimize the effect of nuclear corrections correspondingly.
The resulting data set outlined in Table~\ref{tab:ndp} 
spans the region of $x=5\cdot 10^{-5}\div0.75$ and $Q^2=2.5\div250~$GeV$^2$. 

\begin{figure}
\centerline{\epsfig{file=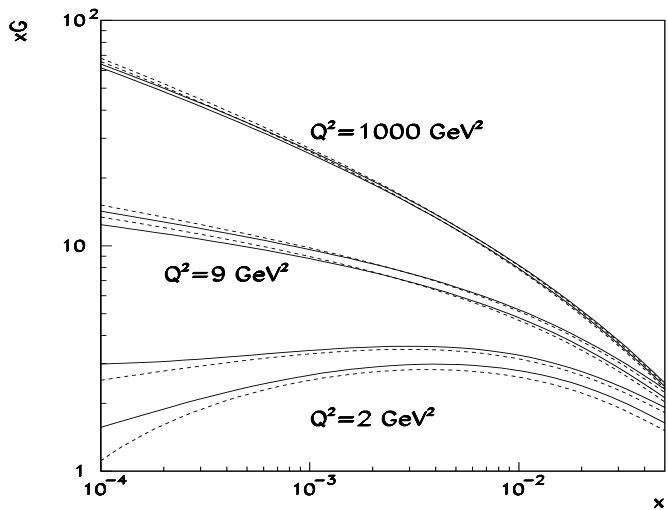,width=10cm,height=8cm}}
\caption{The $1\sigma$ bands for the gluon distributions obtained 
in the NNLO (full lines) and the NLO (dashed lines) analysis
at different values of $Q$.}
\label{fig:pert}
\end{figure}

The statistic and systematic errors in the experimental data  
were combined in the minimized $\chi^2$
using the covariance matrix approach as well as it was done
in our earlier analysis.
The normalization factors for all experiments excluding the old SLAC 
ones were also included into the covariance matrix, while for the 
latter the fitted re-normalization factors were introduced (see
Ref.~\cite{Alekhin:2001ch} for details).
In Table~\ref{tab:ndp} we give the value of $\chi^2$
obtained after the fit and the contributions of each separate experiment 
to its total value. 
One can see that the value of $\chi^2$ reduced to the number of data points
(NDP) is about unity 
for the total data set. The deviation of the values of $\chi^2/{\rm NDP}$ 
off unity for some separate experiments 
can be attributed to the statistical fluctuations in the most cases.
This allows one to conclude 
that in good approximation the data can be described 
by the statistical model with the Gaussian probability 
functions for all errors 
including the systematic ones and hence the errors in the fitted parameters 
are also Gaussian distributed. The obtained values of the fitted parameters 
with their errors are given in Table~\ref{tab:pars}. 

To examine the sensitivity of our results to the specific choice 
of the NNLO splitting functions we repeated the fit with the choice B for 
splitting functions of Ref.~\cite{vanNeerven:2000wp}. 
The results of this fit are compared to the ones of the nominal fit 
in Table~\ref{tab:pars} 
and in Fig.\ref{fig:pdf}. One can see that the difference 
of the PDFs values obtained in the fits with different choices
of the NNLO splitting functions is largest for the gluon distribution at small 
$x$ and anyway does not exceed the experimental uncertainties in the PDFs
through the whole region of $x$ in question.

\begin{table}[h]
\caption{The values of the fitted parameters of the leading twist PDFs
and the strong coupling constant.} 
\begin{center}
\begin{tabular}{ccccc}   
       &     & NLO                      &\multicolumn{2}{c}{NNLO}\\ \cline{4-5}
       &     &                          &(A+B)/2& B \\\hline
Valence &&&& \\
       &$a_u$&$0.709\pm0.027$&$0.726\pm0.025$  &$0.731\pm0.025$\\
       &$b_u$&$3.911\pm0.051$&$4.023\pm0.049$  &$4.016\pm0.049$\\
       &$\gamma_2^u$&$1.06\pm0.35$&$1.04\pm0.33$  &$1.02\pm0.32$\\
       &$a_d$&$0.706\pm0.073$&$0.762\pm0.072$      &$0.792\pm0.071$\\
       &$b_d$&$4.95\pm0.12$&$5.15\pm0.13$         &$5.18\pm0.15$\\
Glue   &&&&\\
       &$a_G$&$-0.145\pm0.019$&$-0.121\pm0.022$   &$-0.082\pm0.022$\\
       &$b_G$&$8.2\pm1.3$&$9.2\pm1.1$     &$9.9\pm1.0$\\
       &$\gamma_1^G$&$-3.79\pm0.45$&$-3.93\pm0.52$   &$-4.37\pm0.45$\\
       &$\gamma_2^G$&$7.7\pm1.7$&$8.4\pm1.7$   &$9.4\pm1.5$\\
Sea    &&&&\\
       &$A_S$&$0.165\pm0.011$&$0.1616\pm0.0091$&$0.1614\pm0.0082$\\
       &$a_{sd}$&$-0.1961\pm0.0048$&$-0.2088\pm0.0044$&$-0.2068\pm0.0042$\\
       &$b_{sd}$&$4.7\pm1.3$&$5.2\pm1.2$&$5.7\pm1.2$\\
       &$\eta_{u}$&$1.16\pm0.11$&$1.13\pm0.10$&$1.09\pm0.10$\\
       &$b_{su}$&$10.42\pm0.86$&$10.72\pm0.84$&$10.57\pm0.83$\\
       &&&&\\
       &$\alpha_s(M_Z)$&$0.1171\pm0.0015$&$0.1143\pm0.0013$&$0.1146\pm0.0012$\\
\end{tabular}
\end{center}
\label{tab:pars}
\end{table}
We also performed fit to the same data within the NLO approximation
in order to check the perturbative stability of our analysis.
The comparison of the results of this fit with the NNLO ones is 
given in Table~\ref{tab:pars}.
One can see that the main difference between the NLO and NNLO 
results is for the value of $\alpha_{\rm s}$ and for the parameters describing 
the sea and the gluon distributions at small $x$. 
Nevertheless, as one can see from  Fig.\ref{fig:pert},
even in this region the shift of the 
NNLO gluon distribution as compared to the NLO one
is of the order of magnitude of their experimental errors
in the wide range of $Q$. The same is valid for the sea distribution 
and other distributions are even less sensitive to the 
inclusion of the NNLO corrections. 
The N$^3$LO QCD corrections as they were estimated
in Ref.~\cite{vanNeerven:2001pe} should have smaller effect than the NNLO ones.
The reasonable conclusion based on these 
considerations is that the perturbative stability of 
the obtained NNLO PDFs is better than their experimental uncertainties
at $x\gtrsim 0.0001$. A particular 
feature of our analysis is that our gluon distribution is positive 
up to $Q\sim 1$~GeV$^2$ in the region of $x\gtrsim0.0001$,
in contrast with the gluon distribution obtained 
in the analysis of Ref.~\cite{Martin:2000gq}.

Our value of 
\begin{equation}
\alpha^{\rm NNLO}_{\rm s}(M_{\rm Z})=0.1143\pm0.0013~({\rm exp)}
\label{eqn:alsa}
\end{equation} 
is by $2\sigma$ lower than the value of 
\begin{equation}
\alpha^{\rm NNLO}_{\rm s}(M_{\rm Z})=0.1166\pm0.0009~({\rm exp)}
\label{eqn:alyn}
\end{equation} 
obtained in the analysis of the similar data set performed in 
Ref.~\cite{Santiago:2001mh}. We should also underline other differences 
of those results with ours. Contrary to the results of 
Ref.~\cite{Santiago:2001mh}, we observe sizeable decrease of 
the $\alpha_{\rm s}$ value under inclusion of the NNLO corrections
(compare $\alpha^{\rm NLO}_{\rm s}(M_{\rm Z})=0.1171\pm0.0015~({\rm exp)}$
in our analysis and  
$\alpha^{\rm NLO}_{\rm s}(M_{\rm Z})=0.1155\pm0.0014~({\rm exp)}$
in Ref.~\cite{Santiago:2001mh}). 
In addition, we do not observe the sharp decrease of the 
error in $\alpha_{\rm s}^{\rm NNLO}$ as compared with the error in 
$\alpha_{\rm s}^{\rm NLO}$. Among the most 
probable explanations of these discrepancies is the  
difference in treatment of the experimental data. For example one cannot  
compare the experimental errors in $\alpha_{\rm s}$ given 
in Eqn.(\ref{eqn:alsa}) and in Eqn.(\ref{eqn:alyn}) since the latter
does not account for the systematic errors in data.
Besides, the analysis of Ref.~\cite{Santiago:2001mh} was performed  
assuming that the contribution of the high-twist terms is zero, while 
we fitted this contribution together with other parameters.  
Evidently, since the high-twist contribution 
to $F_2$ and the value of $\alpha_{\rm s}$ 
are strongly anti-correlated, this may take effect both 
on the central value and the error in $\alpha_{\rm s}$. 
Nevertheless for the comprehensive clarification 
of the differences between our 
results and the ones of Ref.~\cite{Santiago:2001mh} a dedicated 
analysis is needed and we suppose to do it in future
as well as the comparison with the earlier NNLO fit to the 
data on the neutrino DIS structure function $F_3^{\nu N}$ (see 
Ref.~\cite{Kataev:2001kk} and references therein).

{\bf 3.} In summary, we performed the analysis of the existing inclusive 
DIS data within the NNLO QCD approximation. The PDFs and the 
value of strong coupling constant 
$\alpha_{\rm s}(M_{\rm Z})=0.1143\pm0.0013~({\rm exp})$ are obtained. 
The sensitivity of the PDFs to the uncertainty in the value 
of the NNLO corrections to the 
splitting functions is analyzed and it is shown that 
the PDFs errors due to this uncertainty 
is generally less than the experimental uncertainty in PDFs
through the region of $x$ spanned by the existing DIS data.
The obtained set of PDFs may be used to reduce the 
higher order QCD uncertainty in the predictions 
of the cross sections of the hard scattering processes
in the hadron collisions. In particular this may be important 
for reliable estimation of the K-factor for the Higgs boson production
(see Ref.~\cite{Catani:2001es} in this connection). 

{\bf Acknowledgments}
 I am indebted to M.Botje for providing the data tables from ZEUS experiment;
A.Kataev and A.Vogt for stimulating discussions.
The work was supported by the RFBR grant 00-02-17432.

\end{document}